\documentclass[aps,prd,showpacs,nofootinbib,twocolumn,superscriptaddress,preprintnumbers]{revtex4}

\usepackage{amssymb}
\usepackage{amsmath}
\usepackage{epsfig}
\usepackage{hyperref}
\usepackage{breakurl}

\usepackage{undertilde}
\makeatletter
\def\simgt{\mathrel{\lower2.5pt\vbox{\lineskip=0pt\baselineskip=0pt
           \hbox{$>$}\hbox{$\sim$}}}}
\def\simlt{\mathrel{\lower2.5pt\vbox{\lineskip=0pt\baselineskip=0pt
           \hbox{$<$}\hbox{$\sim$}}}}
\makeatother

\newcommand{\be}{\begin{equation}}
\newcommand{\ee}{\end{equation}}
\newcommand{\bea}{\begin{eqnarray}}
\newcommand{\eea}{\end{eqnarray}}
\newcommand{\Eq}[1]{Eq.~(\ref{#1})}

\newcommand{\Sec}[1]{Sec.~\ref{#1}}

\newcommand{\vev}[1]{\langle #1 \rangle}

\newcommand{\MPl}{M_{\rm Pl}}
\newcommand{\LL}{\mathcal{L}}
\newcommand{\OO}{\mathcal{O}}

\newcommand{\sslash}[1]{\ensuremath\raisebox{-0.00cm}{{\small\slash}}\hspace{-0.21cm}#1\/}

\begin{document}

\title{Affleck-Dine Cogenesis}

\author{Clifford Cheung}
\affiliation{Berkeley Center for Theoretical Physics, 
  University of California, Berkeley, CA 94720, USA}
\affiliation{Theoretical Physics Group, 
  Lawrence Berkeley National Laboratory, Berkeley, CA 94720, USA}

\author{Kathryn M. Zurek}
\affiliation{Michigan Center for Theoretical Physics, University of Michigan, Ann Arbor, MI 48109, USA}

\begin{abstract}

We propose a novel framework in which the observed baryon and dark matter abundances are simultaneously generated via the Affleck-Dine mechanism.  In its simplest realization, Affleck-Dine cogenesis is accomplished by a single superpotential operator and its $A$-term counterpart.  These operators explicitly break $B-L$ and $X$, the dark matter number, to the diagonal $B-L+X$.  In the early universe these operators stabilize supersymmetric flat directions carrying non-zero $B-L$ and $X$, and impart the requisite CP violation for asymmetry generation.  Because $B-L+X$ is preserved, the resulting $B-L$ and $X$ asymmetries are equal and opposite, though this precise relation may be relaxed if $B-L$ and $X$ are violated separately by additional operators.   Our dark matter candidate is stabilized by $R$-parity and acquires an asymmetric abundance due to its non-zero $X$ number.  For a dark matter mass of order a few GeV, one naturally obtains the observed ratio of energy densities today, $\Omega_{DM}/\Omega_{B} \sim 5$.
These theories typically predict macroscopic lifetimes for the lightest observable supersymmetric particle as it decays to the dark matter.

\end{abstract}


\maketitle

\section{Introduction}

The existence of the baryon asymmetry and dark matter (DM) are key pieces of evidence for physics beyond the standard model (SM).  In particular, the SM provides neither enough CP violation  to generate the observed baryon asymmetry nor a viable DM candidate.  On the other hand, supersymmetry can accommodate both, albeit through unrelated mechanisms.  The baryon asymmetry is set by new CP violating phases and out of equilibrium dynamics, while the DM density arises from thermal freeze out.

In this paper we unify the production of baryon and DM number through a simple extension of the Affleck-Dine mechanism \cite{AD,ADReview} which exploits the fact that supersymmetric flat directions can also carry DM number.  In particular, we consider a setup with the usual $U(1)_{B-L}$ symmetry carried by MSSM fields and a $U(1)_X$ symmetry carried by additional states which we refer to collectively as the DM sector.  
Typically, there exists an operator
\bea
\OO_{B-L} \OO_{X},
\label{operator}
\eea
where  $\OO_{B-L}$ and $\OO_X$ are gauge invariant products of chiral superfields which carry $B-L$ and $X$ number, respectively.
In general, we are interested in operators of the form 
\bea
\OO_{B-L} &=& LH_u, LLE^c, QLD^c, U^cD^cD^c,
\label{examples}
\eea
which have charge $-1$ under $U(1)_{B-L}$, while we  choose $X$ charges such that $\OO_X$ has  charge $+1$ under $U(1)_X$.   In this convention, $\OO_{B-L}\OO_X$ explicitly breaks $B-L$ and $X$ number down to an exact, diagonal $B-L+X$ number.

As in canonical AD, inflation induces supersymmetry breaking effects proportional to the Hubble parameter which can efficiently drive $\vev{B-L}$ and $\vev{X}$ to non-zero values in the early universe.  As the universe cools, these operators become ineffective and the vacuum settles to the present day $B-L$ and $X$ preserving minimum.  During this transition, the $A$-term counterpart of the operator in \Eq{operator} enters into the scalar potential and induces a ``torque'' on the phases of the complex scalar fields.  This $A$-term provides the required CP violation needed to generate $B-L$ and $X$ asymmetries.  Because the theory preserves $B-L+X$, the resulting  asymmetry has vanishing $B-L+X$ number, so
\bea - n_{B-L} = n_X \neq 0.
\label{conserved}
\eea
 Since the baryon and DM asymmetries are produced simultaneously, we refer to this mechanism as AD ``cogenesis.''  
 The relation in \Eq{conserved} can be modified in the presence of additional operators which separately violate $B-L$ and $X$.

As we will see, the DM sector is thermalized after inflation, albeit at a low temperature, and chemical equilibrium distributes the initial $n_X$ asymmetry among all $X$ charged states which are sufficiently long-lived to freeze out.
An example of such a state is the lightest $X$ number charged particle (LXP), which is often meta-stable, but will in general decay late to $B-L$ charged SM states via $\OO_{B-L}\OO_X$.   In this paper, we will assume that the lightest supersymmetric particle (LSP) carries $X$ number and it thus attains an asymmetric relic abundance from the initial $X$ asymmetry.  Moreover, because the lightest observable supersymmetric particle (LOSP) and the LXP are typically long-lived, this class of theories accommodates an interesting collider phenomenology.

Operators of the form $\OO_{B-L} \OO_X$ were considered more generally in Asymmetric DM \cite{ADM}, which relates a present day asymmetry in baryons and DM via similar symmetry considerations.  However, while in \cite{ADM} the baryon asymmetry was assumed initially and then shared with the DM, in the present work the baryon and DM asymmetries are generated dynamically and simultaneously.  Other types of mechanisms for generating or transferring an asymmetry between sectors have been discussed in the literature, from electroweak sphalerons \cite{sphalerons}, to out of equilibrium decay of heavy particles \cite{outofequilibrium}, and phase transitions in hidden sectors \cite{firstorder}.  Other works on DM with an asymmetry and their phenomenological implications include \cite{otherrefs}. 
A common origin of DM and the baryon asymmetry through the AD mechanism has also been considered via fragmentation of the AD condensate into $Q$-balls \cite{Q1,fujii}, via a sneutrino condensate \cite{sneutrino}.  Finally, we note that during the completion of this work, \cite{Bell} also proposed a model that employs the AD mechanism to produce $B-L$ and $X$ asymmetries.

The outline of this paper is as follows.  In  \Sec{cog} we describe the mechanism of AD cogenesis in general terms.  This will include a discussion of the formation of the AD condensate in the inflationary epoch, as well as its subsequent cosmological evolution after inflation ends.  We then go on in \Sec{post} to discuss the decay of the inflaton and the AD condensate, followed by the ensuing thermal histories of the MSSM and DM sectors.  Afterwards we present a number of simple explicit models of AD cogenesis and their associated variations in \Sec{models}, and discuss the collider phenomenology of these theories in \Sec{pheno}.  Finally, we conclude in \Sec{conclude}.

\section{Cogenesis in the Early Universe}
\label{cog}

Our aim is to simultaneously generate a $B-L$ and $X$ asymmetry at the end of inflation via the evolution of AD condensates which carry $B-L$ and $X$.  
To understand what is required in order to achieve this, let us map our system onto a simple mechanical analog.  In particular, by parameterizing a scalar field $\phi$ in polar coordinates, 
\bea
\phi &=& \frac{1}{\sqrt{2}}r_\phi e^{i \theta_\phi} ,
\eea one finds that the charge density of $\phi$ is
\bea
n_\phi &=& j^0
= i (\phi \dot\phi^\dagger - \phi^\dagger \dot \phi) =  r^2_\phi \dot \theta_\phi,
\label{nphi}
\eea
that is, identical to the angular momentum of a pseudo-particle in two dimensions.   

It is convenient to reinterpret the scalar sector of the MSSM during inflation as a system of coupled pseudo-particles in two dimensions with a time dependent potential. Thus to produce a $B-L$ and $X$ asymmetry we must have a setup in which the initial angular momenta of all the pseudo-particles are vanishing but the final angular momenta in the $B-L$ and $X$ directions are non-zero.  
 Hence, the essential ingredients of our setup are:
\begin{itemize}
\item[i)] {\bf Stabilization.}  Since a torque requires a lever arm, scalar fields must be stabilized away from the origin in the early universe in such a way that both $B-L$ and $X$ are spontaneously broken.
\item[ii)] {\bf Torque.}  For a torque to be exerted, the scalar potential must vary in time and depend explicitly on the phases of fields which are $B-L$ and $X$ covariant. 
\end{itemize}
These criteria are of course equivalent to the Sakharov conditions requiring i) $B-L$ and $X$ symmetry violation and ii) CP violation.  Let us now discuss how each of these elements are accommodated during the formation and evolution of the AD condensate.  

\subsection{Stabilization}

The first phase of the AD mechanism, stabilization, occurs during the initial inflationary epoch of the early universe.  As discussed thoroughly in \cite{Randall1,Randall2}, the expansion of the universe affects the evolution of scalar fields through Hubble friction and through the scalar potential, which takes the form
\bea
V &=& V_F + V_D + V_{\rm soft},
\eea
where $V_F$ and $V_D$ arise from supersymmetric $F$-terms and $D$-terms.  Here $V_{\rm soft}$ will vary explicitly in time via the Hubble parameter because supersymmetry is broken by the vacuum energy of the universe during inflation.  Indeed, Hubble dependent potential terms should be present as a consequence of interactions between the scalar fields and the inflaton induced by Planck scale dynamics.  The presence of these Hubble induced interactions along with Hubble friction implies that the scalar fields are critically damped during the inflationary phase \cite{Randall1,Randall2}.

Typically, $V_{\rm soft}$ will induce additional minima far from the origin.
For example, the AD mechanism exploits the existence of soft mass terms of the form \cite{Randall1,Randall2} 
\bea
V_{\rm soft} &\supset&
\sum_\phi (a_\phi m^2+ b_\phi H^2) |\phi|^2,
\label{tachyon}
\eea
where $m$ is the scale of soft masses at zero temperature and $H$ is the Hubble parameter.  The dimensionless parameters $a_\phi$ and $b_\phi$ are generated by the couplings of the field $\phi$ to the goldstino and the inflaton, respectively.  
In general, it is possible that $b_\phi < 0$ in \Eq{tachyon}, in which case a tachyon is induced for $\phi$ during inflation, causing $\phi$ to roll away from the origin and be stabilized at $\phi$-breaking minimum.  

We should also expect a contribution to the potential from the $A$-term version of $\OO_{B-L}\OO_X$ of the form
\bea
V_{\rm soft} &\supset & (f m + g H) \frac{\OO_{B-L} \OO_X }{ M^{d-4}}.
\label{softtorque}
\eea
where $f$ and $g$ are dimensionless coefficients and $M$ is the scale suppressing the dimension $d$ operator in \Eq{operator}.   As we will see in explicit models in \Sec{models}, this operator introduces additional vacua at non-zero field values. To our knowledge, the possibility that the $A$-term alone, without Hubble tachyons, can drive the AD evolution has not before been pointed out in the literature.
Be it through contributions from \Eq{tachyon} or \Eq{softtorque}, $\phi$ will be naturally pushed along $D$-flat directions until it is lifted by higher order terms in the potential at some large field value.  This state is the AD condensate.

A variety of operators, which may or may not break $B-L$, $X$, or supersymmetry, can serve to lift the flat directions.  For instance, \Eq{operator} is a very natural superpotential operator which is fully supersymmetric, breaks $B-L$ and $X$ down to the diagonal $B-L+X$, and produces a stabilizing $V_F$ potential.  Alternatively, $V_F$ can have stabilizing contributions from supersymmetric operators which separately preserve $B-L$ and $X$.  Also, it is possible that higher order terms from $V_{\rm soft}$ successfully stabilize the field directions.  Finally, we note that additional $D$-terms from a gauged $B-L+X$ symmetry are a particularly elegant way of stabilizing fields with $B-L$ and $X$ number simultaneously.  In \Sec{models} we will explicitly realize some of these stabilizing mechanisms in a number of concrete models.

\subsection{Torque}

Following the inflationary epoch comes the second ingredient of the AD mechanism, torque.  When inflation ends, the universe begins to cool and the energy density is dominated by the coherent oscillations of the inflaton.  During this time, the AD condensate more or less tracks the minimum of the scalar potential, which moves as a function of the Hubble parameter.  If the parameters $f$ and $g$ in \Eq{softtorque} have different phases, then a torque will be exerted on the  phases of the fields in $\OO_{B-L}$ and $\OO_X$ when $H \sim f m/g $.  As the phases of $B-L$ and $X$ evolve from their initial to final values, a non-zero asymmetry in $B-L$ and $X$ develops, as indicated in Eq.~(\ref{nphi}).  

We can now calculate the asymmetry in Eq.~(\ref{nphi}) by tracking the evolution of the scalar fields through the equations of motion for the angular components of $B-L$ and $X$.  
We are interested in the Lagrangian for the angular components of the coupled $B-L$ and $X$ system.  First, we parameterize all fields according to their charges under $B-L$ and $X$, so
\bea
\phi = r_\phi \exp i \left({q_{B-L,\phi} \theta_{B-L}  + q_{X,\phi} \theta_X}\right),
\label{phaseparam}
\eea
where $q_{B-L,\phi}$ and $q_{X,\phi}$ are the $B-L$ and $X$ charges of $\phi$, and $\theta_{B-L}$ and $\theta_X$ are phases which shift by a unit under $B-L$ and $X$, respectively.  In this notation, the Lagrangian is
\bea
{\cal L} &=& \frac{1}{2}(r_{B-L}^2\dot{\theta}_{B-L}^2 + r_X^2\dot{\theta}_X^2) -V (\theta_{B-L}-\theta_X),
\label{tworotors}
\eea
where we have defined the quantities
\bea
 r_{B-L}^2 &=& \sum_{\phi}  q_{B-L,\phi}^2 r_\phi^2 \\
 r_X^2 &=& \sum_\phi q_{X,\phi}^2 r_\phi^2.
 \eea
One can think of $r_{B-L}$ and $r_X$ as the lever arms corresponding to $B-L$ and $X$ number.  In this notation, the $B-L$ and $X$ number densities are
\bea
n_{B-L} &=& r_{B-L}^2 \dot\theta_{B-L} \\
n_X &=& r_X^2 \dot\theta_{X}.
\eea
The parameterization in \Eq{phaseparam} implies that
\bea
\OO_{B-L} &=&  |\OO_{B-L}| e^{-i \theta_{B-L}} \nonumber \\
\OO_{X} & = & |\OO_{X}| e^{i \theta_{X}},
\label{phiop}
\eea
which in turn means that the term in \Eq{softtorque} generates the angular potential shown in \Eq{tworotors}.  As mentioned earlier, $\OO_{B-L}$ and $\OO_X$ have, without loss of generality, been defined to have charge $-1$  under $B-L$ and charge $+1$ under $X$, respectively.  Defining sum and difference angular variables,
\bea
\theta_\pm &=& \theta_{B-L} \pm \theta_X,
\label{newangles}
\eea
we see that the angular Lagrangian has no dependence on $\theta_+$. This implies that conjugate momentum to $\theta_+$, that is the $B-L+X$ number density, is conserved,
\bea
\frac{d}{dt}\frac{\partial {\cal L}}{\partial \dot{\theta}_+}= \frac{d}{dt}\left(n_{B-L} + n_X \right)= 0,
\eea
or equivalently, $B-L+X$ number is conserved at its initial value of zero:
\bea
n_{B-L} + n_X=0.
\label{Lconserved}
\eea
On the other hand, the operator in \Eq{softtorque} explicitly breaks  $B-L-X$, so it generates an effective, time dependent potential for $\theta_-$.  The conjugate momentum, $\partial \LL / \partial \theta_-$, is $B-L-X$ number and is not conserved:
\bea
\frac{d}{dt}\frac{\partial {\cal L}}{\partial \dot{\theta}_-}=\frac{d}{d t} \left(n_{B-L} - n_X\right)=- \frac{\partial V}{\partial {\theta_-}}.
\label{Lnonconserved}
\eea
This equation of motion can be solved parametrically using \Eq{softtorque} and the parameterization in \Eq{phiop}, treating the torque as an impulse occurring at time $H \sim f m/g$.  One finds
\bea
-n_{B-L} = n_X \sim \frac{\arg(f/g) \, g\,  |\OO_{B-L}|\, |\OO_X|}{M^{d-4}},
\label{asymmetry}
\eea
where $|\OO_{B-L}|$ and $|\OO_X|$ are evaluated when $H\sim f m/g$.
Thus, an asymmetry in $B-L$ and $X$ is generated and AD cogenesis is realized.    For the potentials we consider, the AD condensate will typically produce a symmetric abundance of $B-L$ and $X$ charged fields as well.  We will discuss the fate of this symmetric component in \Sec{post} and present a more detailed calculation of the asymmetric component in \Sec{models}, when we consider explicit models. 

Note that the relationship in \Eq{asymmetry} can be modified in the presence of additional operators which separately violate $B-L$ and $X$, such as a Majorana mass term for a field that carries $X$ number.  The presence of the Majorana term, if it is comparable or larger than the soft mass term, can give a significant additional contribution to the $X$ asymmetry which will violate \Eq{asymmetry}.  We will consider this contribution in detail on a case by case basis in \Sec{models}.  

\section{Cosmology after Cogenesis}

\label{post}

Thus far we have established how an initial asymmetry in $B-L$ and $X$ number can be generated via AD cogenesis in the early universe.  It now remains to discuss the effects of inflaton and AD condensate decays on the MSSM and dark sector evolution.  We discuss these aspects next before moving on to specific models.

\subsection{Inflaton Decay}

During AD cogenesis, stabilization and torque are conveniently provided by Hubble induced potential terms generated by the inflaton, which dominates the energy density of the universe as it oscillates towards the origin.  Eventually, however, the inflaton will decay at a reheating temperature $T_R$ defined as the temperature at which the Hubble parameter is equal to the inflaton decay rate.  This subsequently reheats, to some extent, both the MSSM and DM sectors.  This reheating process is highly sensitive to the couplings of the inflaton to the various fields.  For example, one expects  Kahler operators of the form
\bea
K \supset \sum_\phi  \frac{b_\phi}{\MPl^2} \chi^\dagger \chi \phi^\dagger \phi,
\eea
where $\chi$ is the inflaton chiral superfield and $b_\phi$ is the same coefficient fixing the Hubble soft mass of $\phi$ in \Eq{tachyon}.  In this paper we take the natural assumption that $b_\phi$ is comparable for MSSM and DM sector fields, since it is generated by unspecified Planck scale physics.  Thus, the inflaton will decay to both sectors at a similar rate, and both sectors will be comparably reheated.  Relaxing this assumption, especially in cases where the DM sector is reheated very little, leads to interesting cosmological scenarios.  We leave an exploration of these possibilities to future work, and instead focus here on the case where both sectors are reheated equally.  

Naively, an equal degree of reheating into the MSSM and DM sectors has cosmological dangers, given stringent bounds from big bang nucleosynthesis (BBN) constraining the number of light degrees of freedom present at MeV temperatures.   As we will see explicitly in \Sec{thermwash}, however, the two sectors can in general be thermally decoupled from each other immediately after reheating, henceforth evolving to different temperatures.  Indeed, variations in the number of degrees of freedom in the MSSM and DM sectors during the evolution of the universe can substantially alter the relative temperatures of the MSSM and DM sectors \cite{FengHidden}.  Thus, if the DM sector is even modestly cooler than the MSSM during BBN, say even by an order of magnitude in temperature, then these BBN bounds permit many hundreds of degrees of freedom in the DM sector.

Another cosmological pitfall arising from inflaton decays to the MSSM is the overproduction of weakly coupled, stable particles, {\em e.g.}~the gravitino problem \cite{MoroiMurayama} and the axino problem \cite{axinoproblem}.  For example, as is well-known, gravitino overclosure places a bound of at least $T_R \lesssim 10^{10}$ GeV which becomes even more stringent for lower supersymmetry breaking scales.  This is an important constraint on the AD mechanism in general.

Importantly, $T_R$ is also constrained via the observed baryon and DM densities produced in AD cogenesis according to the usual expression for the asymmetric yield \cite{Randall2},
\bea
\eta_B = \frac{n_B}{s} \sim \frac{n_B}{\rho_\chi/T_R},
\label{eta}
\eea
where the inflaton energy density $\rho_\chi$ sets the expansion rate during inflaton dominated reheating, $\rho_\chi \sim H^2 \MPl^2$.   Here $n_B$ and $\rho_\chi$ should be evaluated shortly after AD cogenesis, when Hubble is of order the scale of soft masses.
Because the present day asymmetric yield of baryons is measured to be $\eta_B \sim 10^{-10}$, this relation effectively fixes $T_R$ in terms of the number asymmetry generated by AD cogenesis, which is in turn fixed by the strength of the $\OO_{B-L}\OO_X$ operator.  Lastly, note one final constraint on $T_R$, which is that the Hubble parameter during reheating must be smaller than the scale of soft masses, taken to be of order the weak scale.  If this is not the case, then the inflaton will have decayed too soon to be able to generate the Hubble induced potential terms which drive the AD condensate evolution.  This places a bound of approximately $T_R \lesssim 10^{10} \mbox{ GeV}$.

\subsection{Condensate Decay}

After the initial asymmetry is produced, the universe cools and the AD condensate in-spirals towards the origin, as dictated by the zero temperature scalar potential.  
As discussed in \cite{Q1,Q2,Q3}, if the scalar potential is shallower than quadratic near the origin, then it supports a class of non-topologically stabilized solitons known as $Q$-balls.  If formed, $Q$-balls will be cosmologically stable if their energy density per unit charge is less than that of the lightest $B-L$ or $X$ charged particle.  It has been shown that theories of gauge mediated supersymmetry breaking generally allow for $Q$-ball formation \cite{Q1,Q2}.  On the other hand, whether this occurs in the case of gravity mediation depends sensitively on the precise form of the radiative potential and is thus very model dependent \cite{Q3}.    Throughout this work, we assume a gravity mediated scenario in which the potential does not permit $Q$-ball formation.

In the absence of $Q$-balls, the AD condensate eventually ``evaporates'' as a consequence of scattering with the thermalized decay products of the inflaton.  This evaporation yields symmetric and asymmetric abundances of DM sector particles, with relative sizes determined by the radial and angular velocities of the condensate.  The symmetric component is absorbed by the DM sector bath, but eventually freezes out once the universe sufficiently cools.  In order for AD cogenesis to successfully explain the proximity of the baryon and DM abundances today, the symmetric component of DM must be efficiently annihilated away, leaving a remnant asymmetric relic density.  This is easily accommodated in explicit models, which we consider in greater detail in \Sec{models}. For the present discussion, let us assume that this annihilation occurs efficiently and consider only the asymmetric component.  

Because the DM sector is thermalized at reheating, the $n_X$ asymmetry will be shared among all sufficiently long-lived $X$ carrying particles. Because the $X$ number distribution process is sensitive to the relative $X$ numbers of these states, the precise distribution of the asymmetry is model dependent.  Nevertheless, one finds that the asymmetries are roughly equal
\bea
n_X \sim n_{LXP} \sim n_{LSP},
\eea
up to integer charge factors.  Note that we have assumed that the LSP carries $X$ number, so the proximity of $\Omega_{DM}$ to $\Omega_B$ is explained if $m_{LSP}$ is within an order of magnitude of a GeV.  In this sense, AD cogenesis can address the coincidence problem.  In addition, note that the precise ratio of the DM mass to the proton mass depends on how the baryon or lepton number generated by the AD mechanism is redistributed by the sphalerons to $B$ and $L$.  This in turn depends on details of the electroweak phase transition (EWPT), as described in \cite{HarveyTurner}.

On the other hand, if the LXP has no other stabilizing symmetry, then $n_{LXP}$ will decay back into the SM via $\OO_{B-L} \OO_X$.  In this case the baryon asymmetry will be partially but not completely depleted by the decay, since the LSP carries $X$ number and is completely stable.  The amount of dilution will depend on whether the decay happens before or after the EWPT. It is also possible that the LXP decays so late that it is cosmologically long-lived.   
 For example, if $\OO_{B-L}\OO_X$ is a dimension six, GUT suppressed operator, then the LXP is decaying DM.  The LSP, which also carries $X$ number, comprises an additional component of DM, so in this scenario we have two DM particles, one of which decays.

Finally, let us briefly comment on a viable theory in which the LSP is $X$ neutral, and yet the cosmological evolution still yields the correct DM abundance today.  In particular, assume that the NLSP carries $X$ number and is sufficiently long-lived as to freeze out.  In this case, chemical equilibrium will relate $n_X \sim n_{NLSP}$.  Assuming that the symmetric component of the NLSP is annihilated away, then the asymmetric component will decay to the LSP out of equilibrium.  Hence, the coincidence problem is addressed as long as the LSP mass is of order the GeV scale.  This possibility can be realized by a simple model in which the LSP is a GeV scale gravitino and the NLSP carries $X$ number. Because this theory requires gauge mediated supersymmetry breaking, $Q$-balls typically form out of the AD condensate.  However, if these $Q$-balls only carry $L$ or $X$ number, then they will be unstable and promptly decay to leptons or DM sector particles.

\subsection{Thermalization and Washout}

\label{thermwash}

After the AD condensate and the inflaton decay, the thermal histories of the MSSM and DM sectors begin.
In this section we are interested in addressing two questions about the thermal histories of the MSSM and DM sectors after the decays of the inflaton and the condensate.   First, for which values of $T_R$ will the MSSM and the DM sector be in thermal equilibrium?  Thermalization can occur through a variety of operators which may or may not break $B-L$ and $X$ number.
Second, at what $T_R$ are washout processes efficient?  Washout effects will largely be dictated by when $B-L-X$ violating operators such as $\OO_{B-L}\OO_X$ are in  equilibrium. 

Consider first the scenario in which the MSSM and the DM sector are coupled via an irrelevant operator of dimension $d$ suppressed by a scale $M$.  These interactions decouple at a temperature below
\bea
T_{D}^{(d=5)} &\sim& 10^{14} \textrm{ GeV } \left(\frac{g_*}{200}\right)^{1/2} \left(\frac{M}{10^{15} \textrm{ GeV}}\right)^{2}\\
T_{D}^{(d=6)} &\sim& 10^{14} \textrm{ GeV } \left(\frac{g_*}{200}\right)^{1/6} \left(\frac{M}{10^{15} \textrm{ GeV}}\right)^{4/3} \nonumber.
\label{higherdim}
\eea
Consequently, if $T_{R}$ is below these threshold temperatures, than the associated processes are out of equilibrium.    

In general, operators which connect the MSSM and DM sectors while preserving $B-L$ and $X$ number separately will be $d=6$ and are often the least important.   For instance, this is the case if $B-L+X$ is gauged but spontaneously broken at a high scale $M$, yielding Kahler operators of the form $Q^\dagger Q X^\dagger X /M^2$ at low energies.  The one exception to this statement is the $d=5$ superpotential operator, $H_u H_d X X'$, where $X$ and $X'$ are oppositely charged DM sector states.  
On the other hand, operators coupling the MSSM and DM sectors which break $B-L$ and $X$ number down to the diagonal $B-L+X$ number are often $d=5$, {\em e.g.}~$U^c D^c D^c X$.  Hence, these leading operators can often dictate both the thermalization and washout effects. 
Since, in the presence of $B-L$ and $X$ violation only through $B-L+X$ preserving operators, no net $B-L+X$ asymmetry arises, these $d=5$ operators must be out of equilibrium at the end of inflation to prevent washout of the $B-L$ and $X$ asymmetries.  

The only case in which the operator coupling the MSSM and the DM sector is marginal is $LH_u X$.  In this scenario, associated interactions are decoupled when the temperature is above
\bea
T_{D}^{(d=4)} &=& 100 \textrm{ GeV } \left(\frac{200}{g_*}\right)^{1/2}  \left( \frac{\lambda}{10^{-7}}\right)^2,
\label{margop}
\eea
where $\lambda$ is the associated dimensionless coupling.  As long as the recoupling temperature is below the EWPT, these processes will not wash out the baryon or DM asymmetries.  Furthermore, below the EWPT these processes are kinematically suppressed, so the temperature scaling of interaction rate changes from $\sim T$ to $\sim T^5$, where the latter is the rate for two to two scattering processes.  Thus if the washout processes are out of equilibrium at the weak scale, then they will remain out of equilibrium throughout the history of the universe.   A similar recoupling temperature exists for scattering processes involving a light $B-L+X$ gauge boson, though there is no kinematic suppression so processes can recouple below the EWPT.

Lastly, note that some washout through $X$ violating processes is acceptable, and in this case the DM can be considerably heavier than the GeV scale.  Recalling that our LSP is assumed to carry $X$ number, the final DM abundance will be suppressed with respect to the initial abundance from cogenesis by an amount $ (m_{LSP} /T_{D})^{3/2} \exp{(-m_{LSP}/T_{D})}$, where $T_D$ is the decoupling temperature of the $X$ violating processes.

\section{Explicit Models of Cogenesis}

\label{models}

Next, let us present some explicit models of AD cogenesis.  In particular, we will study models in which $\OO_{B-L} = QLD^c, LH_u$, although most of our statements will apply equally well to any of the theories shown in \Eq{examples}.

\subsection{$QLD^cX$ Operator}
\label{QLDXsec}

Consider an explicit model in which the AD condensate resides on the $QLD^cX$ flat direction.  
The $D$-term potential, $V_D$, arising from the SM gauge group fixes the $D$-flat directions,
\bea
r_Q = r_L = r_{D^c},
\label{QLDXflat}
\eea
in the notation of \Eq{phaseparam}, while $r_X$ is free.
We assume the presence of an $F$-term potential, $V_F$, arising from a superpotential term,
\bea
W &=& \frac{QLD^c X } {M},
\label{QLDXsuper}
\eea
as well as its $A$-term partner,
\bea
V_{\rm soft} &=&   
(f m + g H) \frac{QLD^cX}{M}.
\label{QLDXsoft}
\eea
In general, there will be zero temperature soft masses of order $m$, but they will not play an important role in the AD evolution other than to ensure that the origin is a stable minimum at late times, so we neglect these terms.  Furthermore, one should typically expect soft masses induced by Hubble expansion, which are usually assumed to be tachyonic in canonical AD constructions.  For the following analysis we ignore such contributions to \Eq{QLDXsoft} in order to showcase the fact that the $A$-term alone can produce induce vacuum away from the origin.  We will discuss the effects of Hubble induced soft masses later on.

Plugging in \Eq{phaseparam} and \Eq{QLDXflat} into the full scalar potential yields
\bea
V &=&
\frac{r_{Q}^6}{8M^2} + \frac{3 r_{Q}^4 r_X^2}{8M^2} \nonumber \\
&& + \frac{r_{Q}^3 r_X}{2M} f m \cos(\arg f - \theta_{B-L}+\theta_X) \nonumber \\
&&+ \frac{r_{Q}^3 r_X}{2M} g H \cos(\arg g - \theta_{B-L}+\theta_X) .
\label{QLDXpot}
\eea
At early times the second term can be ignored because it is proportional to $m$.  The angular components naturally align to make the cosine term in the third line negative, and then the potential is stabilized by the supersymmetric terms in the first line.  We find the potential has an extremum at
\bea
&& r_{Q}^2  = r_X^2 = \frac{2g HM}{3} \nonumber \\
&& \arg g - \theta_{B-L} + \theta_X = \pi,
\label{QLDXvevs}
\eea
where one can check easily that this extremum is stable.  Hence, an AD condensate can form at this point in the early universe.

As the universe cools, eventually $H\sim  f m /g$, and a torque is applied to the condensate by the cosine term in \Eq{QLDXpot}.  Plugging \Eq{QLDXvevs} into \Eq{asymmetry}, we obtain an estimate for the asymmetry given by
\bea
-n_{B-L} = n_X \sim \arg(f/g) \, f^2 g\, m^2 M.
\label{QLDXasymmetry}
\eea
This result agrees with numerical simulations to within an order of magnitude.  After the $B-L$ and $X$ asymmetries are produced, the AD condensate then evolves and eventually decays to the DM particle, as per the general discussion given in \Sec{post}.

According to \Eq{eta}, the asymmetric yield can be expressed in terms of the number density in \Eq{QLDXasymmetry}, $\rho_\chi$, and $T_R$.  Demanding that $\eta_B \sim 10^{-10}$ thus fixes $T_R$ as a function of $M$.  At the same time, the usual constraints from gravitino overclosure require the conservative bound, $T_R \lesssim 10^{10}$ GeV.  Putting it all together, given order one values for $f$ and $g$, one finds a bound of approximately $M \gtrsim 10^{16}$ GeV.  Interestingly, $M$ is required to be near or above the GUT scale.

Such a high cutoff introduces some tension with BBN bounds. In particular, assuming that the LOSP decays into the DM sector solely through \Eq{QLDXsuper}, then the associated lifetimes will be quite long.  These decays will typically produce electromagnetic and hadronic energy which can destroy the successful predictions of BBN.  As is well known, however, these BBN bounds are contingent on the nature and freeze out abundances of the LOSP, which are highly model dependent.  Moreover, there can easily exist additional higher dimension operators on top of \Eq{QLDXsuper} which are suppressed by a lower cutoff and mediate a faster decay of the LOSP into the DM sector.  These additional operators can separately preserve $B-L$ and $X$ number in such a way that the evolution of the AD condensate will be more or less unaltered from the discussion above.

The $QLD^cX$ model described above is extremely simple because it simultaneously stabilizes and exerts a torque on the AD condensate using only the operators in \Eq{QLDXsuper} and \Eq{QLDXsoft}.  That said, this minimal  model accommodates a number of interesting variations.

First of all, one can add additional operators beyond those shown in \Eq{QLDXsuper} and \Eq{QLDXsoft}.  Hubble induced soft masses of the form in \Eq{tachyon} are in general present, and they will influence the AD evolution because they are parametrically comparable in strength to the torque term in \Eq{QLDXsoft}. Irrespective of whether these soft masses are tachyonic or not, they can alter the numerical coefficients in \Eq{QLDXvevs} and \Eq{QLDXasymmetry}, leaving the parametric dependences unchanged.  In addition, since $B-L+X$ number is exact in this model, it is very natural to gauge this symmetry.  The associated $D$-term potential then imposes an additional stabilization constraint on the fields beyond \Eq{QLDXflat}, given by $r_Q^2 = r_X^2$.  Hence, gauging $B-L+X$ is a very natural mechanism for simultaneously fixing both $B-L$ and $X$ number to non-zero values in the early universe.

Secondly, variations of this model exist with additional DM sector particles which are charged under $U(1)_X$.  In the early universe, these additional states may be stabilized at the origin or not.   Indeed, as long as the $X$ field is stabilized away from the origin then AD cogenesis is accommodated.  Additional DM sector states can serve a number of purposes, for instance providing the fermionic component of $X$ a Dirac mass via $m_D X X'$.  Note that a mass for $X$ is not a requirement.  As discussed earlier, there naively exists stringent bounds from BBN on additional light or massless degrees of freedom, but these are easily sidestepped if the DM sector is thermally decoupled from and modestly cooler than the MSSM bath during BBN \cite{FengHidden}.

As noted earlier, because the DM sector is thermalized there will in general be a symmetric abundance of DM particles in the DM sector bath.  Removing this symmetric component requires the existence of additional interactions, which require additional $X$ carrying states.  For instance, symmetric annihilation is accomplished using a Yukawa coupling $\kappa X {X'}^2$ for sufficiently large $\kappa$.  Alternatively, one has the option of introducing additional gauge bosons in the DM sector.

\subsection{$LH_u X$ Operator}

Next, consider a model in which the AD condensate resides on the $LH_u X$ flat direction.  The mechanics of this theory are largely similar to those of the $QLD^c X$ operator.  In this case, the $D$-flat directions fix
\bea
r_L = r_{H_u},
\label{LHXflat}
\eea
where we use the notation of \Eq{phaseparam}, and here $r_X$ is a priori unconstrained.  This model is defined by the superpotential
\bea
W &=& \lambda LH_u X ,
\label{LHXsuper}
\eea
and the analogous $A$-term,
\bea
V_{\rm soft} &=&  
(f m + g H) \lambda L H_u X,
\label{LHXsoft}
\eea
where $\lambda$ is a dimensionless coupling which is much less than unity.
As before, we ignore zero temperature soft masses of order $m$. 
The full scalar potential is given by
\bea
V &=& \frac{\lambda^2 r_L^4}{4} + \frac{\lambda^2 r_L^2 r_X^2}{2} \nonumber \\
&& +\frac{\lambda r_L^2 r_X f m \cos(\arg f - \theta_{B-L} + \theta_X)}{\sqrt{2}} \nonumber  \\
&&+ \frac{\lambda r_L^2 r_X g H \cos(\arg g - \theta_{B-L} + \theta_X)}{\sqrt{2}}.
\label{LHXpot}
\eea
The angular variables align to make the cosine term negative, and the runaway direction is stabilized by the supersymmetric terms, yielding a minimum at
\bea
&& r_{L}^2  = r_X^2 = \frac{g^2 H^2}{2 \lambda^2} \nonumber \\
&& \arg g - \theta_{B-L} + \theta_X = \pi.
\label{LHXvevs}
\eea
Note that the AD condensate is stabilized further from the origin for smaller values of $\lambda$.  
When eventually $H\sim  f m /g$, the cosine term in \Eq{LHXpot} yields an asymmetry, estimated in general in \Eq{asymmetry}, given by
\bea
-n_{B-L} = n_X \sim \frac{\arg(f/g) \, f^3 g\, m^3}{4\lambda^2},
\eea
which accords with numerical simulations.  The asymmetric yield today is given by \Eq{eta}, which, fixing $\eta_B \sim 10^{-10}$, implies a constraint on $T_R$ in terms of the small coupling $\lambda$.  Combining this with the bound from gravitino overproduction, $T_R < 10^{10}$ GeV, we find that the coupling constant must be less than $\lambda \lesssim 10^{-8}$ in this theory assuming order one values for $f$ and $g$.    Unlike in the $QLD^cX$ theory, this coupling is sufficiently large that the $LH_uX$ model does not in general suffer from the BBN problem of late LOSP decays into the DM sector.

Since no net $B-L+X$ asymmetry is generated, there is also the constraint that the $L H_u X$ operator does not wash out the $B-L$ and $X$ asymmetries.  As computed in \Eq{margop}, interactions in the thermal plasma involving this operator place a bound of $\lambda \lesssim 10^{-7}$.  

As in the $QLD^c X$ model, the $LH_uX$ model has many variations, depending on whether additional operators or fields are added.  
However, for the $LH_uX$ model there is an additional complication, which is that $L$ and $X$ mix after electroweak symmetry breaking.  Consequently, the couplings of $X$ are closely connected and thus constrained by neutrino physics.  There are a number of ways of accommodating the measured active neutrino masses with the presence of the operator $LH_u X$.  For instance, one can simply fix $\lambda \sim 10^{-12}$, yielding Dirac neutrino masses in the eV range.   
Alternatively, one can add a Dirac mass term, $m_D X X'$, which at low energies leaves the active neutrino sector completely unaffected since it exactly preserves $B-L$ and $X$ number.

Lastly, consider the case that $X$ has a Majorana mass term, $m_M X^2$.   Here we imagine that $m_M$ ranges from an eV up to a TeV.  Because the Majorana mass violates $X$ number explicitly, it will affect the evolution of the AD condensate so that \Eq{conserved} is not exactly true.  Moreover, there will be scattering processes in the DM sector bath that include a Majorana mass insertion and tend to wash out the $X$ asymmetry.  Concretely, consider interactions involving the Yukawa coupling $\kappa X X'^2$ suggested in \Sec{QLDXsec}.  We are interested in a process involving  $\kappa$ as well as the insertion of a factor of $m_M$, the leading spurion for $X$ number breaking.  At temperatures far above the mass of the $X$ particle,  any process involving this Majorana mass insertion has a cross-section suppressed by a factor of $(m_M /T)^2$.  Hence, washout effects decouple when the DM sector is at temperatures above $T_{\sslash{X}}$ where
\bea
&& \xi^{-2/3}T_{\sslash{X}} \sim \frac{ m_M^{2/3} \kappa^{4/3} \MPl^{1/3}}{g_*^{1/6} } \\
&&\sim  10^5 \textrm{ GeV} \; \left( \frac{\kappa}{0.1}\right)^{4/3} \left( \frac{m}{50 \textrm{ GeV}}\right)^{2/3}  \left( \frac{200}{g_*}\right)^{1/6},\nonumber
\label{washcouple}
\eea
and where $\xi$ is the ratio of the DM sector temperature to the MSSM sector temperature.
In order to save the asymmetry, we require that  $T_{\sslash{X}} \lesssim m_M/20$, the freeze out temperature of $X$, which cannot be satisfied for any reasonable value for $m_M$.  Hence, it is difficult to accommodate the usual seesaw origins of the active neutrino masses in this framework of annihilation to DM sector states.

On the other hand, the annihilation may occur through SM states, such as the $Z$ boson.  If this is the case, then washout is suppressed by insertions of $\lambda$ provided that $\lambda\lesssim 10^{-7}$, so the associated processes become inefficient at the weak scale.  In order to generate the the eV neutrino mass scale, the Majorana mass for the DM must be GeV scale.  The scenario with this set of parameters was explored in \cite{MarchRussell}.

\section{Collider Signatures}

\label{pheno}

In this section we outline possible collider signatures associated with models of AD cogenesis.  As we will see, the phenomenology is largely dictated by the structure of the connector operator $\OO_{B-L}\OO_X$ and so the models typically have a degeneracy with other models which employ this portal.  

We have assumed throughout that the LSP carries $X$ number,  so it resides in the DM sector.  Consequently, supersymmetric collider phenomenology is drastically altered, since the LOSP necessarily decays into the DM sector due to $R$-parity conservation.  In the minimal scenario, this decay is mediated by $\OO_{B-L}\OO_X$.  As we saw in \Sec{models}, the coefficient of this operator is bounded collectively from gravitino overproduction and the observed relic abundance of baryons and DM.   

For the $QLD^cX$ model, and more generally for any model with $\OO_{B-L}\OO_X$ dimension five, these constraints imply that $M \gtrsim 10^{16}$ GeV.  Thus, the decay of, {\em e.g.}~a squark LOSP via $\tilde q \rightarrow \ell q \tilde x$, will be long-lived on collider time scales.  While naively problematic, the associated collider signatures can be quite spectacular if the LOSP is charged or colored.  In this case some fraction of LOSPs produced will ionize and eventually stop within the detector material, then decay late and out of time with the beam.  A number of proposals exist to measure these stopped LOSP decays \cite{Stopper,StoppingGluinos}, and indeed, CMS has already performed a search of this kind \cite{CMS}.

In contrast, consider the $LH_uX$ model.  As we saw in \Sec{models}, the coupling constant is bounded by $\lambda \lesssim 10^{-8}$.  Thus, the decay length of a chargino LOSP decaying via $\tilde C \rightarrow \ell \tilde x$ is
\bea
c\tau &\sim&  1 \mbox{ cm}\times \left(\frac{100 \mbox{ GeV}}{m}\right)\left(\frac{10^{-8}}{\lambda}\right)^2,
\eea
ignoring mixing angles.  Hence, the LOSP is typically displaced, and in some cases even long-lived.
Remarkably, if $\lambda \sim 10^{-12}$, as is necessary for Dirac neutrino masses, then the LOSP is stable on detector time scales.  See \cite{CliffSeesaw} for a detailed study of $LH_u X$ and its effect on supersymmetric collider phenomenology and neutrino physics.

\section{Conclusions}
\label{conclude}

In this paper we have proposed a unified framework for baryon and DM number generation using a simple extension of the AD mechanism.   Our setup exploits the  possibility that supersymmetric flat directions can carry both $B-L$ and $X$ number.  The asymmetries are generated by operators of the form $\OO_{B-L}\OO_X$ and their CP violating $A$-term counterparts.  Indeed, the very same $A$-terms which provide the CP violating torque also aid in stabilizing the $B-L$ and $X$ number carrying fields away from the origin.  
Because the relevant interactions separately violate $B-L$ and $X$ but preserve $B-L+X$, equal and opposite $X$ and $B-L$ asymmetries are produced.  Thus, AD cogenesis naturally addresses the coincidence of $\Omega_{DM}/\Omega_B\sim5$ if the LSP carries $X$ number and has a mass of order the GeV scale.  

The collider phenomenology of these models is quite remarkable because the LOSP will decay to the LSP via $\OO_{B-L}\OO_X$, the very same operator responsible for the asymmetry generation.  As we have shown, this operator is required to be quite weak in order to avoid washout and accommodate the observed relic abundances today.  Thus, the LOSP is typically displaced or long-lived on the time scales of collider physics, allowing for distinctive signatures from stopped meta-stable charged particles.

While the explicit models presented in this paper are purposefully minimal, they offer a fertile starting point from which to understand the full space of possibilities and complications of AD cogenesis theories.   Still, there remain important aspects of cogenesis which warrant detailed future study, for instance a comprehensive analysis of the formation and stability of the AD condensate and its subsequent decay to particles.  Likewise, a systematic understanding of the viable cosmological histories within this framework is left to future work.

\begin{acknowledgements}
We thank Hai-Bo Yu for collaboration during the early stages of this work.
C.C.~is supported in part by the Director, Office of Science, Office of High Energy and Nuclear Physics, of the US Department of Energy under Contract DE-AC02-05CH11231 and by the National Science Foundation on grant PHY-0457315.

\end{acknowledgements}

\end{document}